\documentclass[prl,showpacs,superscriptaddress,twocolumn]{revtex4-1}

\usepackage{graphicx, subfigure}

\usepackage{amsmath}
\usepackage{amsfonts}
\usepackage{graphicx}

\usepackage{natbib}
\usepackage[final=true]{hyperref}

\usepackage{color}
\usepackage[usenames,dvipsnames]{xcolor}

\usepackage{xcolor}
\usepackage[dvipsnames]{xcolor}

\newcommand{\rbb}{(\mathbf{r})}

\begin{document}
\title{Stationary quantum vortex street in a driven-dissipative quantum fluid of light}

\author{S.~V.~Koniakhin}
\email{kon@mail.ioffe.ru}
\affiliation{Institut Pascal, PHOTON-N2, Universit\'e Clermont Auvergne, CNRS, SIGMA Clermont, Institut Pascal, F-63000 Clermont-Ferrand, France}
\affiliation{St. Petersburg Academic University - Nanotechnology Research and Education Centre of the Russian Academy of Sciences, 194021 St. Petersburg, Russia}

\author{O. Bleu}
\affiliation{Institut Pascal, PHOTON-N2, Universit\'e Clermont Auvergne, CNRS, SIGMA Clermont, Institut Pascal, F-63000 Clermont-Ferrand, France}
\affiliation{ARC Centre of Excellence in Future Low-Energy Electronics Technologies and School of Physics and Astronomy, Monash University, Melbourne, VIC 3800, Australia}

\author{D.~D.~Stupin}
\affiliation{St. Petersburg Academic University - Nanotechnology Research and Education Centre of the Russian Academy of Sciences, 194021 St. Petersburg, Russia}

\author{S. Pigeon}
\affiliation{Laboratoire Kastler Brossel, Sorbonne Universit\'e, CNRS,ENS-PSL Research University, College de France, 4 place Jussieu, 75252 Paris, France}
\author{A. Maitre}
\affiliation{Laboratoire Kastler Brossel, Sorbonne Universit\'e, CNRS,ENS-PSL Research University, College de France, 4 place Jussieu, 75252 Paris, France}
\author{F. Claude}
\affiliation{Laboratoire Kastler Brossel, Sorbonne Universit\'e, CNRS,ENS-PSL Research University, College de France, 4 place Jussieu, 75252 Paris, France}
\author{G.~Lerario}
\affiliation{Laboratoire Kastler Brossel, Sorbonne Universit\'e, CNRS,ENS-PSL Research University, College de France, 4 place Jussieu, 75252 Paris, France}
\affiliation{CNR NANOTEC, Istituto di Nanotecnologia, via Monteroni, 73100 Lecce, Italy}
\author{Q. Glorieux}
\affiliation{Laboratoire Kastler Brossel, Sorbonne Universit\'e, CNRS,ENS-PSL Research University, College de France, 4 place Jussieu, 75252 Paris, France}
\author{A. Bramati}
\affiliation{Laboratoire Kastler Brossel, Sorbonne Universit\'e, CNRS,ENS-PSL Research University, College de France, 4 place Jussieu, 75252 Paris, France}

\author{D.~Solnyshkov}
\affiliation{Institut Pascal, PHOTON-N2, Universit\'e Clermont Auvergne, CNRS, SIGMA Clermont, Institut Pascal, F-63000 Clermont-Ferrand, France}

\author{G.~Malpuech}
\affiliation{Institut Pascal, PHOTON-N2, Universit\'e Clermont Auvergne, CNRS, SIGMA Clermont, Institut Pascal, F-63000 Clermont-Ferrand, France}

\begin{abstract}
We investigate the formation of a new class of density-phase defects in a resonantly driven 2D quantum fluid of light. The system bistability allows the formation of low density regions containing  density-phase singularities confined between high density regions. We show that in 1D channels, an odd (1-3) or even (2-4) number of dark solitons form parallel to the channel axis in order to accommodate the phase constraint induced by the pumps in the barriers. These soliton molecules are typically unstable and evolve toward stationary symmetric or anti-symmetric arrays of vortex streets straightforwardly observable in \emph{cw} experiments. The flexibility of this photonic platform allows implementing more complicated potentials such as maze-like channels, with the vortex streets connecting the entrances and thus solving the maze.
\end{abstract}

\maketitle

A fluid is called quantum when it exhibits quantum-mechanical effects at a macroscopic scale. For bosons, it occurs when many particles can be described by a single-particle wave function. This collective behaviour can arise spontaneously when particles undergo a phase transition towards a quantum-coherent state such as a superconducting state, a superfluid, or a Bose-Einstein condensate (BEC). Cavity exciton-polaritons \cite{kavokin2} (polaritons) are 2D photonic modes interacting via their excitonic parts. Their quantum coherence can spontaneously occur through a BEC process \cite{kasprzak2006bose}, but a unique feature of this photonic system is that such coherence can also be imprinted by a resonant laser and be preserved substantially longer than their lifetime \cite{wertz2010spontaneous}. 
The high control of the injected flow combined with a direct optical access to the full wave function (amplitude, phase, space, time) make this platform very attractive to study quantum fluid physics \cite{RevModPhys.85.299}. A typical example which revealed the potential of this system is the observation of oblique dark solitons  \cite{pigeon2011Hydrodynamic,amo2011polariton,Hivet}, which form when a supersonic quantum fluid hits a defect (initially proposed in 2006 \cite{PhysRevLett.97.180405} for atomic BECs). The 2D solitons forming behind the defect remain stable because the transverse "snake instability"\cite{KIVSHAR2000117,Dutton2001}, making 2D solitons normally unstable, is carried away by the supersonic flow making the soliton effectively 1D \cite{PhysRevLett.100.160402,Walker2017}. Such supersonic flow is energetically unstable, but polariton flows can be efficiently decoupled from thermal relaxation \cite{wertz2010spontaneous,Wertz2012}, which made possible the observation of oblique dark solitons \cite{amo2011polariton,Hivet}.

An interesting regime occurs if the fluid velocity is decreased just below the speed of sound. In such a case, the subsonic flow still interacts with the defect exhibiting a local acceleration.
This leads to the formation of quantum vortex streets composed of vortex-antivortex pairs. 
This quantum version of von Karman vortex streets can be understood as the  decay of the oblique solitons via the snake instability when the stabilization by the supersonic flow is lots.
 The creation of vortex-antivortex pairs has been reported in  time-resolved pulsed experiments, both in polaritons \cite{sanvitto2011all,nardin2011hydrodynamic} and atomic quantum fluids \cite{Kwon2016}. Similar proposals were made for non-resonantly pumped polaritons \cite{Smirnov2014,Liew2015}. However, the study of the snake instability dynamics leading to quantum vortex streets requires both \emph{cw} excitation and time resolution, and it remained elusive so far. In a recent theoretical work, it was proposed to improve this scheme by sustaining the propagating flow against radiative decay by using a support laser covering the whole sample \cite{pigeon2017sustained}. Interestingly, this configuration demonstrates original density-phase defects.
This pump-support scheme is not limited to the study of the flow scattering on defects, but can be used in a more general frame to create and study a large variety of topological defects \cite{Sich2012,Tercas2014,Sich2018}. Topological defects in the driven-dissipative case can be stationary only when the support laser intensity falls in the bistability loop of the non-linear system \cite{Baas2004Opticalbistability,Yulin2008,Zhang2012,pigeon2017sustained}, where the density can be either low or high, depending on the laser absorption.  Stationary phase defects exist in low-density regions, where the phase is not fixed, because most particles are not directly injected by the laser but diffuse from higher-density regions. The control of the spatial distribution of intensity and phase allows to realize various confining potentials, such as 1D channels, 0D traps \cite{PhysRevB.89.134501}, or circuits made by the combination of both.

In this work, we show that the creation of narrow 1D low-density channels surrounded by high-density regions leads to the formation of stationary dark soliton molecules. These molecules exhibit snake instabilities, leading to the formation of coupled chains of vortex-antivortex pairs -- stationary vortex streets, stabilized by the confining potential. We finally show that the pump distribution can be used to create a maze. Vortex streets retreat from the dead ends, while the entrance and the exit of the maze remain connected, implementing an efficient analog all-optical maze solving algorithm.

The resonantly pumped microcavity is modelled by the standard driven-dissipative Gross-Pitaevskii equation, formally equivalent to the Lugiato-Lefever equation \cite{Lugiato1987}. We neglect the polarization degree of freedom, the non-parabolicity of the polariton dispersion, and any thermal effects \cite{Vishnevsky2012,Stepanov2018}. The equation reads:
\begin{equation}
i\hbar\frac{\partial \psi }{\partial t}  = \left[-\frac{\hbar^2\nabla
^2}{2m} -i \Gamma + g\left| \psi\right|^2 \right]\psi + (S+P)e^{-i \omega_0 t},
\label{eq_GPEph}
\end{equation}
where $\Gamma =\hbar/(2 \tau)$ is the polariton decay rate ($\tau=15$~ps), $m=8\times 10^{-5}m_0$ is the polariton mass ($m_0$ is the free electron mass), $g=5~\mu$eV$\mu$m$^2$ is the polariton-polariton interaction constant. The detuning between the ground state and the pump laser is $\omega_0=0.14~\mathrm{meV}/\hbar$. The support $S$ and the pump $P(\mathbf{r})$ are at normal incidence.

\begin{figure}[tbp]
  \centering
   \includegraphics[width=1\linewidth]{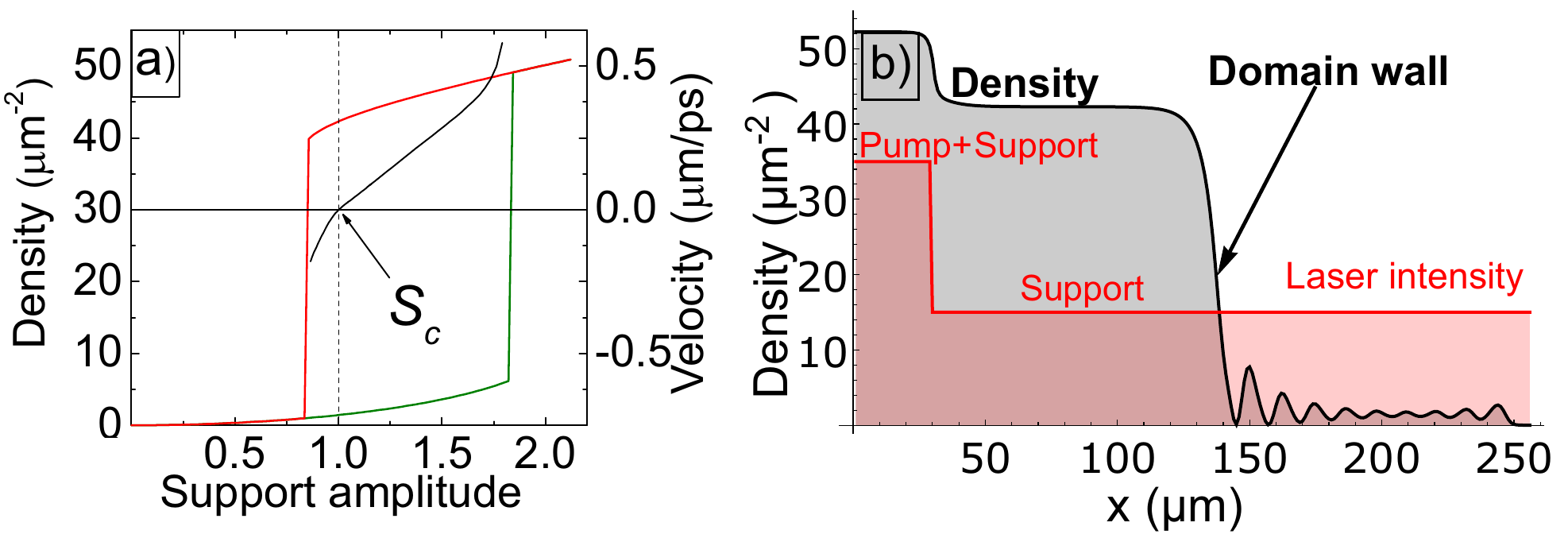}
      \caption{ a) Polariton bistability: density as a function of the pump  (red, green); DW velocity (black, right Y axis).  b) A single high-density region generated by a pump (left), with support present everywhere: condensate density (black curve) combined with the profile of laser intensity (red curve).} \label{fig_wallbist}
\end{figure}

We first consider spatially homogeneous pumping (support only). The bistability loop obtained for our parameters is shown in Fig.~\ref{fig_wallbist}(a). Next, we add a half-space pump ($x<30$~$\mu$m) switching the system to the higher branch of the bistability loop, whereas the other half-space remains on the lower branch. The use of a spatially inhomogeneous pump $P(\mathbf{r})$ allows to control the pumping intensity in the high- and low-density regions independently.

Figure~\ref{fig_wallbist}(b) shows the two regions separated by a domain wall (DW). The region under the pump shows a large intensity and a fixed phase. The intensity at the DW decays within one healing length $\xi=\hbar/\sqrt{2gn m}$ ($n$ is the density) and then exhibits small periodic oscillations of intensity. The DW is stable against the development of instabilities along $Y$, but can propagate along $X$. Such DW propagation has been previously considered for polaritons \cite{kavokin2008neurons,Amo2010Light} and in general for switching waves in optics \cite{Rozanov1997,Ganne2001,odent2016optical} and beyond.
The velocity $v$ of the DW computed numerically as a function of support intensity is shown in Fig.~\ref{fig_wallbist}(a) (black). As expected for this class of differential equations, $v$ is linear in $S-S_c$, where the critical value of the support $S_c$ is given, in the limit $\Gamma\to 0$, by the Maxwell construction \cite{Maxwell1875,suppl}
\begin{equation}
    S_c\approx\frac{2\left(\hbar\omega_0\right)^{3/2}}{3\sqrt{3}g^{1/2}}
\end{equation}
When the support is larger than $S_c$, the DW propagates to the right with a velocity
\begin{equation}
\label{speed}
v \sim 2\frac{S-S_c}{S_c}\frac{\xi}{\tau},
\end{equation}
where $\xi\approx 1.8\mu$m, and the high intensity region expands to the whole space. For support values below $S_c$ the high-density region shrinks and the DW stops at the boundary of the pump ($x=30\mu$m). Around $S_c$, spatially localized solutions of the Gross-Pitaevskii equation bifurcate under the form of dark solitons multiplets \cite{Parra2016}.

\begin{figure*}[tbp]
  \centering
 \includegraphics[width=0.95\linewidth]{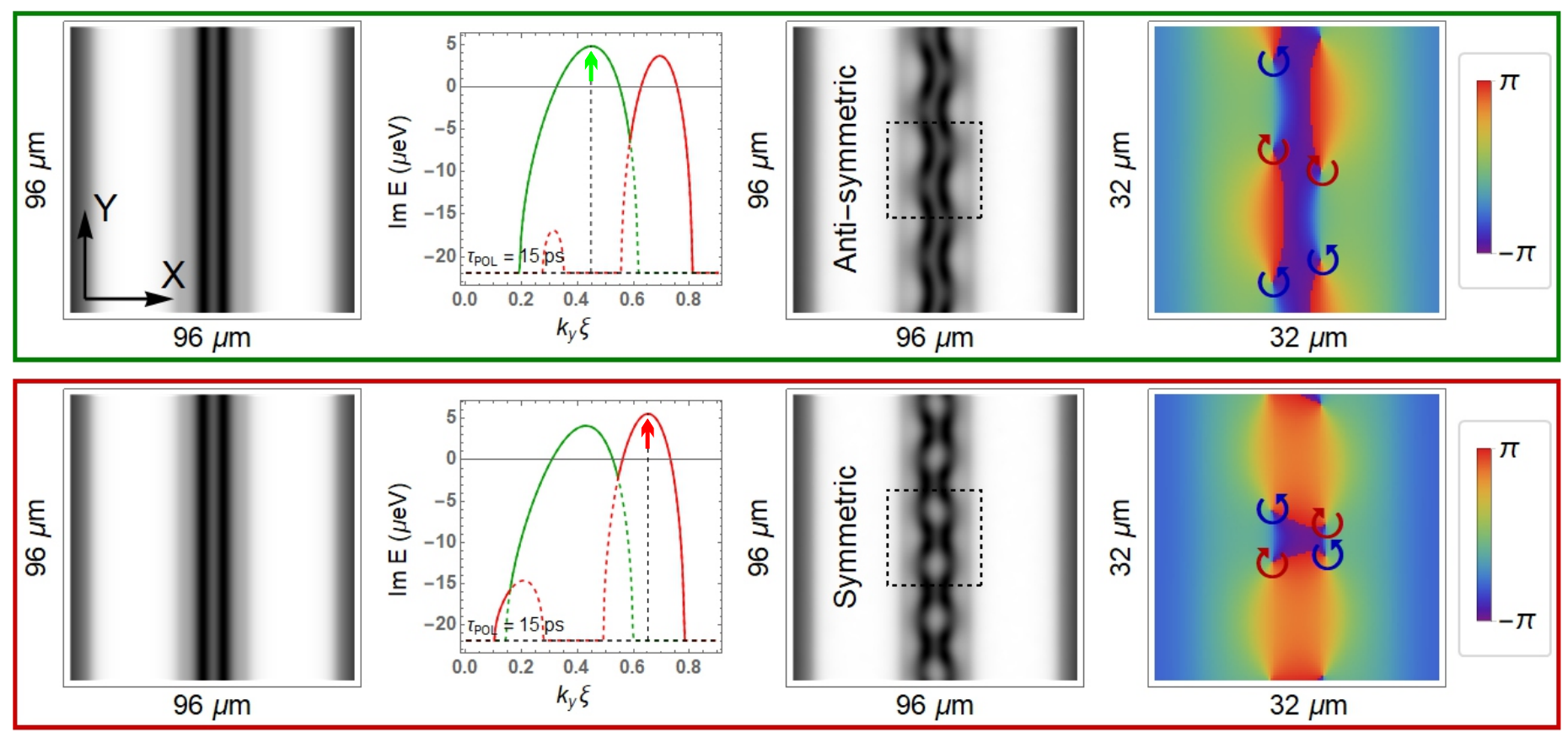}
  \caption{Modulational instability of guided solitons. $L=25$~$\mu$m. $S=0.25S_c$, $P=1.25, 2 S_c$ (top, bottom). Columns: 1) stationary solution with even number of solitons between the high-density walls. 2) imaginary part of the energy of weak excitations of the stationary solution from the 1st column as a function of $k_y$. 3,4): stationary solution after the development of the modulational instability (density, phase)  in the presence of weak disorder. Frame colors correspond to the color of points in Fig.~\ref{fig_dp}.} \label{fig_img}
\end{figure*}

Next, we consider a second high-density region with its boundary parallel to the first one, defining an all-optically controlled confining potential. Similar configuration, but without the support beam, has been studied in~\cite{Aioi2013}.
We start by considering high density regions with the same phase and a fixed channel width $L=23$~$\mu$m  ($\sim 13\xi$ of the high density region). Fig.~\ref{fig_img} is computed at $S=0.25S_c$ for 2 values of $P$. The first column presents the stationary intensity distribution with 2 dark solitons in the channel. The system is effectively 1D, since it is translationally invariant along $Y$ (periodic boundary conditions). Dark solitons are anti-symmetric states with a $\pi$ phase shift. The phase constraints imposed by the high-density regions therefore only allow an even number of solitons. These soliton multiplets are however unstable with respect to the development of instability along $Y$ for a large range of parameters. The second column of Fig.~\ref{fig_img} shows the imaginary part of the energy of the weak excitations versus their longitudinal wave vector $k_y$, obtained from the Bogoliubov-de Gennes equations:
\begin{eqnarray}
\label{BdG}
    L\rbb u\rbb+g\psi\rbb^2v\rbb& =& \hbar \omega u\rbb,\\
    L\rbb v\rbb+g(\psi^*\rbb)^2u\rbb &=& - \hbar \omega v\rbb,\nonumber
\end{eqnarray}
where $L\rbb=-\hbar^2\nabla^2/2m + 2g|\psi\rbb|^2 - \hbar \omega_0 - i\Gamma$. Comparing with Ref. \cite{morgan2013stability}, the chemical potential $\mu$ for conservative system is replaced by the laser frequency in the driven-dissipative case \cite{carusotto2004probing,solnyshkov2008dispersion}. Eqs.~\eqref{BdG} represent an eigenvalue problem for $\omega$. The translational invariance along $Y$ allows to replace $
    - \nabla^2 \rightarrow k_y^2 - \partial^2/\partial x^2,$
where $k_y$ is the wave vector of perturbations.  Positive imaginary part of the energy leads to the development of modulational instability  (the snake instability, well known in conservative condensates \cite{Kuznetsov1988}). The maximal instability wave vector can be estimated \cite{suppl} as $k_y^*=1/(\sqrt{2}\xi)$.
The lowest energy mode in the double-well potential formed by the soliton pair is a symmetric bound state, whereas the highest mode is an anti-symmetric anti-bound state. However, the symmetry of the patterns in the 3rd column of Fig.~\ref{fig_img} is inversed, because each soliton is anti-symmetric by itself ($\pi$ phase jump). An antisymmetric superposition of solitons is therefore a symmetric function, which gives rise to the symmetric pattern observed in the 2nd line of Fig.~\ref{fig_img}. A more detailed study of the mode competition is shown in \cite{suppl}, section II. The mode with a higher imaginary part develops faster (marked by arrows in Fig.~\ref{fig_img}). The 2D modulational instabilities in the numerical simulations can be triggered by any noise or fluctuations breaking the translational symmetry along the $Y$-axis. Here, we consider a weak Gaussian disorder with a correlation length of $2~\mu$m and an amplitude $\gamma=0.01$~meV.  The third and fourth columns of Fig.~\ref{fig_img} show the intensity and the phase of the stationary wave function after the development of the instability. The precise realization of the disorder determines the positioning of the pattern along $Y$, but does not affect the shape, at least if the disorder amplitude is sufficiently small: $\gamma\ll\hbar\omega_0$. Additional simulations shown in \cite{suppl} confirm that these stationary patterns are accessible considering realistic disorder parameters up to 0.1 meV. In all cases, the solitons break into two vortex anti-vortex chains, which can be seen as stationary vortex streets. 2D analysis confirms the stability of the final patterns. It means that the snake instability develops, but is then frozen by the presence of the confining potential. 

\begin{figure}[tpb]
  \centering
  \includegraphics[width=0.9\linewidth]{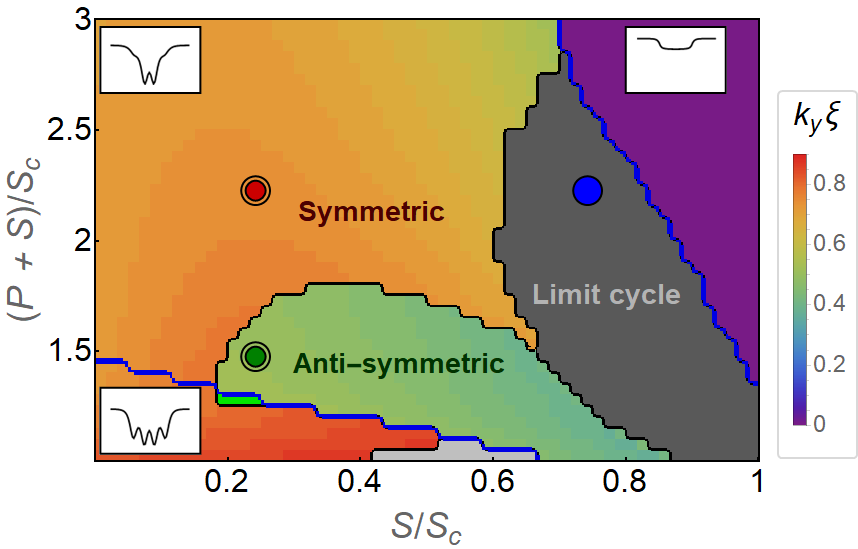}
  \caption{Phase diagram versus support $S$ and pump $P$. The color shows the maximal instability wave vector $k_y^*$. Green tones are for long period antisymmetric excitations (snakes) and orange/red tones are for symmetric excitations with shorter period. Lower left corner separated by the blue curve corresponds to the 4-soliton initial state. Dark gray area is for oscillating in time solitons and violet is for high density in channels (no solitons). Colored dots correspond to the panels in the boxes of the same color in Fig.~2. Blue dot is for maze pathfinding regime (Fig.~4). The insets show the transverse profiles of unperturbed density in the channel.
  } \label{fig_dp}
\end{figure}

Figure~\ref{fig_dp} shows a phase diagram obtained from the stability analysis versus pump and support intensities (constant $L$ and $\hbar\omega_0$). The phase in the upper right corner (purple) corresponds to high density in the channel (no solitons or vortices). It occurs for support values typically smaller than $S_c$, because of the particle flow coming from the two DWs instead of one. This regime is qualitatively similar with the one of the polariton neuron picture \cite{kavokin2008neurons}. The dark grey region corresponds to a non-stationary steady state (limit cycle), at least in the conditions of our simulations, namely without energy relaxation and for sufficiently low disorder. This phase shows a pair of breathing solitons oscillating in time (see movie \cite{suppl}). The small light-grey domain corresponds to a lattice of four solitons. This occurs for small $P$ and large $S$, so weak transverse flows which favours the soliton lattice stability. The next phase located at the bottom left corner corresponds to the collapse of 4 solitons into a symmetric pair of vortex chains (see Fig.~S1 \cite{suppl}). The two next phases located above the blue line correspond to the collapse of  2 solitons into symmetric and anti-symmetric vortex chains respectively (as in Fig.~\ref{fig_img}(a,b)). A tiny domain (lime-green) exhibits the collapse of 4 solitons into an anti-symmetric pair of vortex chains. The false color scale of the figure shows the maximal instability wave vector  $k_y^*$ (except for the non-stationary and stable phases). The anti-symmetric solutions have a twice larger period than the symmetric ones. The $k_y^*$ gradient within a given phase is relatively small, which means that the patterns visibility should not be strongly affected by pump/support intensity fluctuations in a real experiment. Disorder broadens the transitions between the phases, but the core regions remain well defined (see \cite{suppl} for details).

\begin{figure}[tbp]
  \centering
   \includegraphics[width=0.999\linewidth]{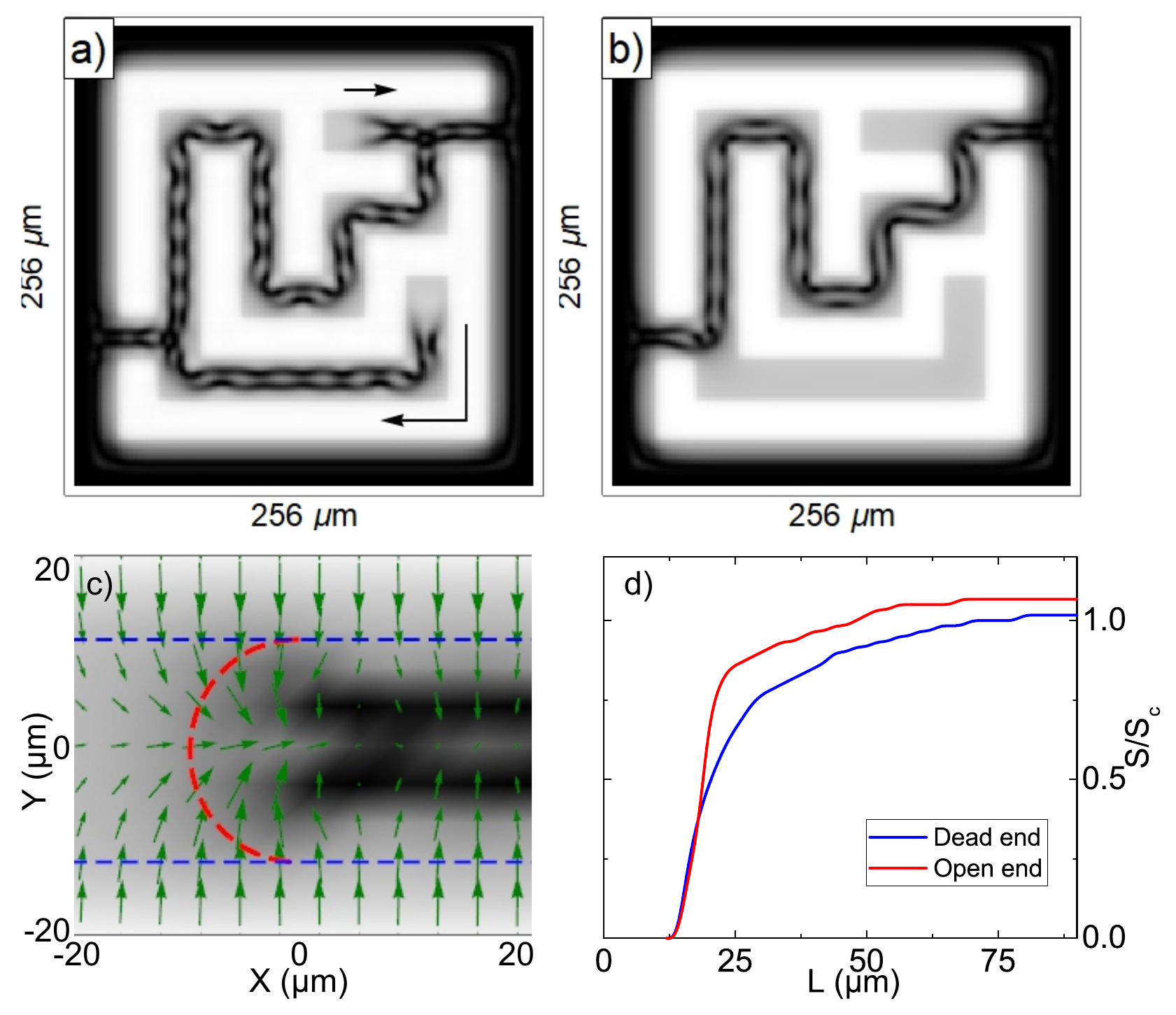}
   \caption{Maze solving: a) initial moments: soliton heads repelled from dead ends; b) shows the stationary final distribution (maze solved); c) DW repelled from the dead end: confinement and particle flows; d) Support threshold vs channel width for dead end (blue) and open end (red).} \label{fig_maze}
\end{figure}

Another interesting possibility offered by this driven-dissipative system is to tune the relative phase of the pump between the walls, working at zero support. In such a case, an odd number of solitons forms  \cite{suppl}, decaying into the same number of chains of  vortex-antivortex pairs. This tuning of the soliton number by varying the relative phase between the pumps is a generalization of  \cite{Goblot2016} to 2D. Here, the 2D character of the system allows instabilities along $Y$. In fact, modern optical techniques allow creating any shape of confining potential, such as various 0D traps \cite{PhysRevB.89.134501}, or graphs, connecting 0D sites with 1D channels. 

The geometry we address now is a maze of 1D channels [Fig.~\ref{fig_maze}(a,b)]. For a proper value of $(S,P)$, immediately after the jump of the walls on the upper branch, the maze is filled with solitons. However, the dead ends represent a configuration different from that of Fig.~\ref{fig_img}: the heads of the vortex streets start to withdraw. 
Fig ~\ref{fig_maze}(a) shows the intensity distribution 20~ps after driving pump and support are switched on (blue circle in Fig.~\ref{fig_dp} in the non-stationary phase). The heads are moving as symbolized by the arrows. Fig ~\ref{fig_maze}(b) shows the final intensity distribution ($t=1$~ns), where the street only connects the two exits of the maze. 
A zoom on the vortex street head is shown in Fig.~\ref{fig_maze}(c).
This head is a DW, but the conditions for its motion are different from the lateral DW in an open 1D channel we considered previously. Indeed, the motion of the head-DW along $X$ is facilitated by the confinement and the flow along $Y$ coming from the two lateral walls. When the head-DW arrives at a cross-roads, an open 1D channel geometry is restored and the DW stops. Fig.~\ref{fig_maze}(d) shows the critical support intensity for the  motion of the two types of DW (dead/open ends -- blue/red). Both decrease when the corridor becomes narrower, tending to zero for $L\approx 14$~$\mu$m. This is the minimal width of a finite quantum well (of $0.14$~meV depth, given by the laser detuning), where the two-node state can exist for such particle mass. For a wide open channel (red), $S$ can even exceed $S_c$, because the quantum pressure of the multinode structure prevents the DWs from meeting each other (see \cite{suppl} for details).
As expected, there is a substantial support range where the head-DW propagates, whereas an open 1D corridor remains stable. Within this range, this configuration represents an optical maze solver (see also a supplementary movie for the dynamics \cite{suppl}).  
The head-DW motion can be affected by disorder, but being an extended object, it is less sensitive than vortices which easily pin on defects \cite{Lagoudakis2009}. Numerical simulations \cite{suppl} show that disorder trapping can be avoided by working at higher detunings $\hbar\omega_0>3\gamma$ and, if necessary, shorter lifetimes, making the effects observable in realistic systems. 

All-optical maze solving is important for the large interdisciplinary field of analog graph solving algorithms \cite{Shannon1951,Steinbock1995,Caruso2016,Berloff2017}. The solving time is determined by the velocity $v$ of the head-DW (see \cite{suppl} for details). 
In the worst case, the length of the dead end is $N L$, where $N$ is the number of cells in the maze (or vertices in the graph) and $L$ is the width of a channel. If $Z$ is the overall system size, the maximal number of cells is $N=Z^2/L^2$. The solving time is therefore $t=NL/v$.
The best among the other maze solving (pathfinding) algorithms \cite{Even2011} such as the Depth-First Search also exhibit the worst-case complexity of $O(N)$, but the practical advantage of the present analog implementation is the small value of the prefactor $L/v\sim 0.5$~ns: the high velocity $v$ reduces the solving time, allowing such an analog maze solver to outperform a modern PC which needs hundreds of clock ticks to check a single cell.

To conclude, we have shown that the modulational instability can be controlled and stabilized in a driven-dissipative polariton system allowing the on-demand formation of soliton molecules and vortex streets exhibiting a particularly rich phase diagram. Non-stationary regimes can be used for  fast analog maze solving.

\begin{acknowledgments}
We acknowledge the support of the ANR "Quantum Fluids of Light" project (ANR-16-CE30-0021) and of the ANR program "Investissements d'Avenir" through the IDEX-ISITE initiative 16-IDEX-0001 (CAP 20-25). S.V.K. and D.D.S. acknowledge the support from the Ministry of Education and Science of Russian Federation (Project 16.9790.2019). D.S., A.B. and Q.G. acknowledge the support of IUF (Institut Universitaire de France).
This work has received funding from the European Union’s Horizon 2020 research and innovation programme under grant agreement No. 820392 (PhoQuS).
\end{acknowledgments}

\bibliography{bib}

\section{Supplemental Materials}
In this Supplemental Material, we provide additional details on the results shown in the main text. Section I discusses the behavior of isolated domain walls. Section II is devoted to the stability of soliton structures in channels. Section III provides additional results for channels with $\pi$ phase difference. Section IV discusses the behavior of solitons in a maze. Section V describes the two supplementary movies.

\section{Stationary and propagating domain walls}

The behavior of bistable systems and, in particular, spatially inhomogeneous solutions present in such systems were a subject of studies for a long time. An overview on optical systems can be found in Ref. \cite{Rozanov1997}.

We start by looking for a stationary spatially inhomogeneous solution where a part of the system is at the upper bistability branch, whereas another part is at the lower bistability branch. These two parts are separated by a domain wall (or switching wall). This problem can be solved analytically for negligibly small dissipation $\Gamma$ (as compared with the laser detuning $\delta=\hbar\omega_0$), and then the solution can be generalized to non-zero $\Gamma$.

The stationary driven-dissipative Gross-Pitaevskii equation that needs to be satisfied by the wave function reads:
\begin{equation}
\left( { -\hbar\omega_0-\frac{\hbar^2}{2m}\Delta  + g {{\left| {{\psi _0}} \right|}^2} - 
i\Gamma_0 } \right){\psi _0} + S = 0
\label{bistabeq}
\end{equation}
where we are going to neglect the term $\Gamma$ at first. Once it is neglected, the wave function $\psi$ and the pumping (support) $S$ can be assumed to take only real values without loss of generality (since all coefficients are real). The equation~\eqref{bistabeq} can therefore be rewritten as
\begin{equation}
m_0\frac{d^2x}{dt^2}=F(x)
\end{equation}
which is a Newton's equation of motion for a material point ($x=\psi$) with a mass $m_0=\hbar^2/2m$ under the effect of a position-dependent "force"
\begin{equation}
    F(x)=g x^3-\delta x+S
\end{equation}
to which one may attribute a "potential"
\begin{equation}
    U(x)=-\int F(x)~dx=-\frac{g x^4}{4}+\frac{\delta x^2}{2}- Sx
\end{equation}

The two maxima of this potential located at the coordinates $x_1$ and $x_3$ correspond to the two stable domains (high density and low density), while the minimum located at $x_2$ corresponds to the inaccessible part of the bistability curve. The system is stationary only if the values of the effective potentials at the two maxima are exactly the same:
\begin{equation}
\label{maxwell}
    U(x_1)=U(x_3),
\end{equation}
otherwise the domain wall starts to propagate. Indeed, a material point should start its motion at one maximum and finish at the other maximum, and for this the two maxima have to be at the same potential height. The points $x_1$ and $x_3$, corresponding to the extrema of $U(x)$, can be found analytically from the cubic equation $F(x)=0$, and the condition~\eqref{maxwell} can be rewritten as
\begin{equation}
    \int\limits_{x_1}^{x_3}F(x)~dx=0
\end{equation}
corresponding to the Maxwell construction in thermodynamics \cite{Maxwell1875}. Solving this equation analytically for the unknown $S$ gives finally:
\begin{equation}
    S_c=\frac{2\delta^{3/2}}{3\sqrt{3}g^{1/2}}
\end{equation}
and the ratio with respect to the pumping required for bistability $S_{min}=\Gamma_0\sqrt{\hbar\omega/g}$ is given by
\begin{equation}
    \frac{S_c}{S_{min}}=\frac{2}{3\sqrt{3}}\frac{\delta}{\Gamma}
\end{equation}
These analytical expressions presented in the main text can be used as estimates for the required support level. A more precise result can be obtained only numerically, as discussed below.

For non-negligible $\Gamma$, all terms in the equation~\eqref{bistabeq} become comparable. The real terms (the kinetic and the interaction energy, the detuning) are of the order of $\delta=\hbar\omega_0$, while the only imaginary term is $\Gamma$ (imaginary part of kinetic energy is small with respect to $\Gamma$). Thus, the expression for the critical pumping $S_c$ at $\Gamma$ comparable with $\hbar \omega_0$ can be sought by replacing of the first three terms in \eqref{bistabeq} by $\hbar \omega_0$ with some coefficient. This coefficient can be found from numerical simulations by the small variation of parameters $\omega_0 \rightarrow \omega_0 + \Delta\omega_0$, $\Gamma \rightarrow \Gamma + \Delta\Gamma$, and $g \rightarrow g + \Delta g$, which allows obtaining the coefficients ($c_{\omega} \approx 0.75$, $c_{\Gamma} \approx 0.75$, $c_{g} = -0.5$) in the Taylor expansion:

\begin{multline}
    \frac{S_c(\omega_0 + \Delta\omega_0,\Gamma + \Delta\Gamma, g + \Delta g)}{S_c(\omega_0,\Gamma, g)} = 1 + \\ c_{\omega} \frac{\Delta\omega_0}{\omega_0} + c_{\Gamma} \frac{\Delta\Gamma}{\Gamma} + c_{g} \frac{\Delta g}{g}.
    \label{eqTaylor}
\end{multline}

As a net result, one obtains    

\begin{equation}
    S_c=\sqrt{\frac{\hbar\omega_0}{g}}\sqrt{\left(\frac{3\hbar\omega_0}{32}\right)^2+\Gamma^2}
\end{equation}

One sees that this equation has the same Taylor expansion as Eq.~\eqref{eqTaylor} for actual values of system parameters.


For $S\neq S_c$, one of the domains becomes more favorable than the other, and the domain wall starts to propagate. Assuming that the density changes linearly with position across the domain wall (which is valid in the vicinity of the inflection point of this wall), we find the following expression for the speed of the wall:
\begin{equation}
v=\frac{\partial n}{\partial t}\frac{\Delta x}{\Delta n}
\end{equation}
where $\Delta x$ is the width of the domain wall and $\Delta n$ is the difference of the densities in the high and low density regions.

To find the derivative $\partial n/\partial t$, we define $n_0$ as the density at the inflection point which corresponds to $S_c$. The only contribution into $\partial \psi/\partial t$ at $S\neq S_c$ can come from the difference $S-S_c$ (other terms in the time-dependent driven-dissipative Gross-Pitaevskii equation cancel):

\begin{equation}
\frac{\partial \psi}{\partial t}=\frac{S-S_c}{i\hbar}
\end{equation}
which allows to write the solution
\begin{equation}
\psi\left(t\right)=\frac{S-S_c}{i\hbar}t+\sqrt{n_0}e^{i\phi_0}
\end{equation}

This expression strongly depends on the phase $\phi_0$ of the wave function at the inflection point (with respect to the phase of the pump). In a homogeneous system, the phase is given by $\tan\phi=-\Gamma/(g n - \hbar\omega_0)$. In the low-density region, $\phi\to 0$, whereas in the high-density region $\phi\to\pi/2$ just above the threshold density $g n \approx\hbar\omega_0$. We assume that $\phi_0$ takes an intermediate value $\phi_0=\pi/4$, and we can also assume that it changes linearly with $S-S_c$ for small deviations from $S_c$, in which case the time derivative of the density can be found as:
\begin{equation}
\frac{\partial n}{\partial t}=\frac{\partial}{\partial t}\left|\psi\left(t\right)\right|^2\approx 2\sqrt{n_0}\frac{S-S_c}{\hbar}\frac{1}{\sqrt{2}}\left(1-\chi\frac{S-S_c}{S_c}\right)
\end{equation}
For $S\approx S_c$, the second-order correction can be neglected, but it starts to become important for larger deviation of the pumping, creating an important difference between the speed of the wall in the case of shrinking or expanding high-density domain. 

In the simplest case, the speed can be found as
\begin{equation}
\label{speed}
v=\sqrt{2}\sqrt{n_0}\frac{S-S_c}{\hbar}\frac{\Delta x}{\Delta n}
\end{equation}

By estimating $\sqrt{\Delta n} \approx \sqrt{n_0} \approx S_c/\Gamma$ and taking $\Delta x=\sqrt{2}\xi$ one rewrites Eq. \eqref{speed} as 
\begin{equation}
\label{speed}
v \sim 2 \frac{S-S_c}{S_c}\frac{\xi}{\tau}
\end{equation}

The expression \eqref{speed} for the values used in numerical simulations gives $v=12$~$\mu$m/ns, very close to the numerical value of 30~$\mu$m/ns for $(S-S_c)/S_c=0.05$.

\section{Stability of driven-dissipative solitons in 2D}

We analyze the soliton stability in the 2D driven-dissipative configuration using the Bogoliubov-de Gennes equations for weak excitations of the condensate. Let $\psi$ be the non-trivial solitonic solution of the driven-dissipative Gross-Pitaevskii equation written on the polariton basis
\begin{equation}
i\hbar \frac{{\partial {\psi}}}{{\partial t}} =  - \frac{{{\hbar ^2}}}{{2m}}\Delta \psi  + g {\left| \psi  \right|^2}\psi  - i\Gamma \psi  + P(x,y)e^{-i\omega_0 t}
\label{drgpe}
\end{equation}
We can write $\psi=\psi_0(x,y)e^{-i\omega_0 t}$, and the perturbed solution is $\psi_0(x,y)+A(x)e^{i(k_yy-\omega t)}+B^*(x)e^{-i(k_yy-\omega t)}$, where $\omega$ is the perturbation frequency relative to the laser frequency $\omega_0$ and $\psi_0(x,y)$ is the solution of the stationary equation
\begin{equation}
\hbar\omega_0\psi_0=-\frac{\hbar^2}{2m}\Delta\psi_0+g|\psi_0|^2\psi_0-i\Gamma\psi_0+P    
\end{equation}
Linearizing~\eqref{drgpe}, we obtain the Bogoliubov-de Gennes equations for $A$ and $B$:
\begin{widetext}
\begin{equation}\left( {\begin{array}{*{20}{c}}
{ - \frac{{{\hbar ^2}}}{{2m}}\left( {\frac{{{\partial ^2}}}{{\partial {x^2}}} - k_y^2} \right) + 2g {{\left| {{\psi _0}} \right|}^2} - i\Gamma  - \hbar {\omega_0} - \hbar \omega }&{g \psi _0^2}\\
{g \psi _0^{*2}}&{ - \frac{{{\hbar ^2}}}{{2m}}\left( {\frac{{{\partial ^2}}}{{\partial {x^2}}} - k_y^2} \right) + 2g {{\left| {{\psi _0}} \right|}^2} - i\Gamma  - \hbar {\omega_0} + \hbar \omega }
\end{array}} \right)\left( {\begin{array}{*{20}{c}}
A\\
B
\end{array}} \right) = \left( {\begin{array}{*{20}{c}}
0\\
0
\end{array}} \right)
\label{bdg1}
\end{equation}
\end{widetext}
where we are interested in the bogolon states confined along the $X$ direction. 

\subsection{Maximal instability wave vector for a single soliton}

For a single state quantized in a potential trap formed by the profile of a single dark soliton, its energy (which is not yet the bogolon frequency) can be estimated as \cite{Pitaevskii}:
\begin{equation}
    -\frac{\hbar^2}{2m}\frac{\partial^2 A}{\partial x^2}+g|\psi_0|^2A-\hbar\omega_0 A\approx -\frac{\hbar\omega_0}{2}A
\end{equation}
which supposes that the system is above the bistability threshold and that the soliton is almost  dark. The diagonalization of the matrix~\eqref{bdg1} thus gives an equation
\begin{equation}
{\left( {\frac{{{\hbar ^2}k_y^2}}{{2m}} + g {{\left| {{\psi _0}} \right|}^2} - i\Gamma  - \frac{{\hbar {\omega_0}}}{2}} \right)^2} - {g ^2}{\left| {{\psi _0}} \right|^4} = {\hbar ^2}{\omega ^2} 
\label{seclr}
\end{equation}
which allows to estimate the value of the imaginary part of $\omega$. Indeed, if we suppose $\Gamma\ll \hbar\omega_0$, the solution for $\omega$ becomes imaginary if the real part of \eqref{seclr} is negative. Maximal imaginary part is achieved if the first square is zero and the second square is maximized, which is obtained at a point $x_0$ where $g|\psi_0(x_0)|^2=\hbar\omega_0/2$, determining the maximal possible positive imaginary part of $\hbar\omega$ as
\begin{equation}
    \Gamma_{max}=\frac{\hbar\omega_0}{2}-\Gamma
\end{equation}
In this case, the soliton is clearly always unstable, because the negative contribution to the imaginary part is much smaller than the positive one, as required for the bistability. We can conclude that 2D solitons in the driven-dissipative configuration remain unstable with respect to small perturbations, at least if they are obtained at the upper bistability branch.

This also provides a criterion for the maximal wavevector $k_y$ of the unstable region:
\begin{equation}
    k_y^*=\sqrt{\frac{m\omega_0}{\hbar}}
\end{equation}
which is allowed at a different point $x_1$, where $g|\psi_0(x_1)|^2\approx 0$. At the upper part of the bistability curve, where $\hbar\omega_0=gn$, this expression can be linked with the inverse of the healing length of the quantum fluid $\xi$:
\begin{equation}
    k_y^*=\frac{1}{\xi\sqrt{2}}
\end{equation}
This expression is given in the main text.

\subsection{Mode competition}

If several quantized states are present in the trap with the energies $E_n$, the highest energy state forms the bogolon with the highest imaginary part. Indeed, the imaginary part of $\omega$ from the simplified equation
\begin{equation}
    \hbar^2\omega^2=\left(g|\psi_0(x)|^2-E_n\right)^2-g^2|\psi_0(x)|^4
\end{equation}
is due to the second term, which is negative. The minimal value of the first term is zero. It is achieved when $g|\psi_0(x)|^2=E_n$, which determines the value of $g|\psi_0(x)|^2$, and the imaginary part is therefore simply 
\begin{equation}
 \Im(\hbar\omega)=E_n   
\end{equation}
measured from the energy of the bare states. Our conclusion on the mode competition in the trap is therefore that the highest mode in the trap always wins (developing first the instability), as soon as it is sufficiently well localized in it. A more exact result can only be obtained by numerical solution. In numerical analysis, we start by finding the stationary solution of the GPE $\psi_0(x)$, which is then plugged into \eqref{bdg1}. The imaginary parts of the bogolon energy $\hbar\omega$ found by this procedure are plotted in Fig.~2 of the main text, and the winning mode is shown with an arrow.

\begin{figure*}[tbp]
  \centering
  \includegraphics[width=1.0\textwidth]{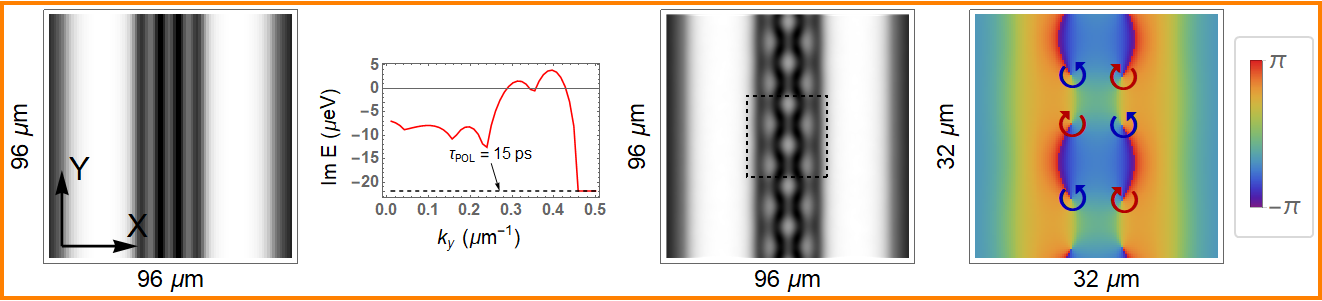}
  \caption{Modulational instability of guided solitons. The phase is the same in the high density walls separated by 25~$\mu$m. $S=0.25S_c$, $P=S_c$. 1st column: stationary solution with even number of solitons between the high-density walls. 2nd column: imaginary part of the energy of weak excitations of the stationary solution from the 1st column as a function of $k_y$. 3rd and 4th column: stationary solution after the development of the modulational instability (density, phase)  in the presence of weak disorder. The color of the frame matches with the colors of the corresponding point in Fig.~3 of main text.} \label{fig_imgS}
\end{figure*}

Figure~\ref{fig_imgS} shows an additional panel with the same elements as Fig.~2 of the main text, corresponding to the lowest pumping case of the phase diagram in Fig.~3 of the main text (orange point), when 4 solitons are formed in the stationary unperturbed regime. These 4 solitons, however, decay into 2 vortex streets once the modulational instability develops.

\subsection{Effects of disorder}

In experiments, the patterns arising from the development of the instability are affected in space by the pinning induced by the disorder \cite{PhysRevB.88.201303}. 

\begin{figure*}
  \centering
  \includegraphics[width=0.8\linewidth]{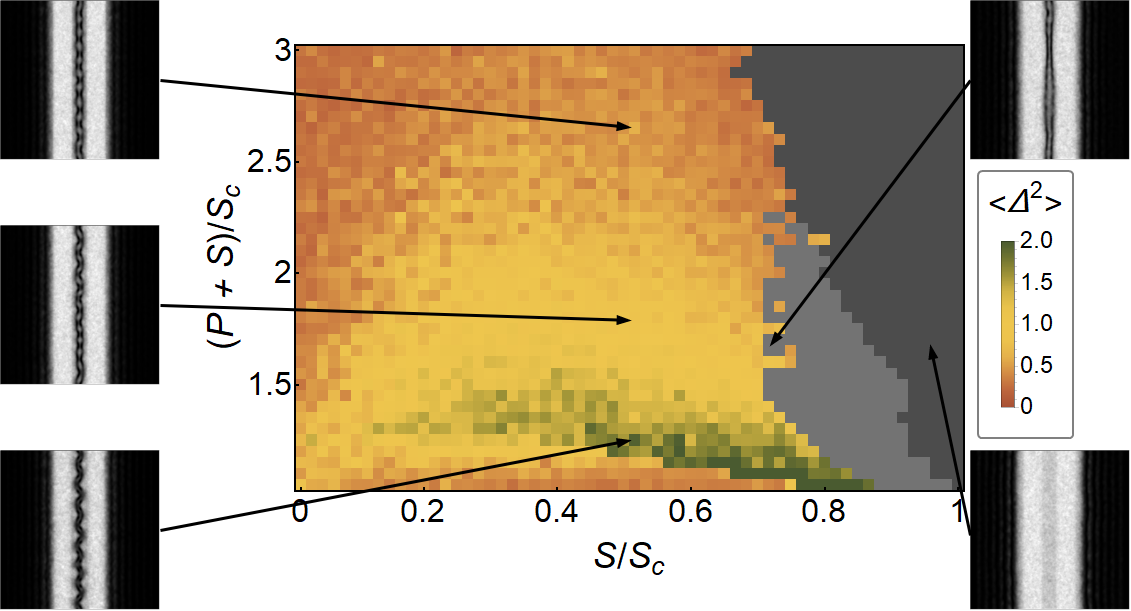}
  \caption{Phase diagram obtained by simulations with disorder. Color shows the standard deviation of center between the solitons $X_c(y)$. Light gray area corresponds to limit cycle phase with no solution stationary in time. Dark gray shows the no-soliton phase with the full filling of the channel with polaritons.} \label{figpdd}
\end{figure*}

Large disorder should lead to a smoothing of the phase diagram. To test the robustness of the phases discussed in the main text with respect to the disorder, we have plotted an alternative phase diagram (Fig.~\ref{figpdd}) where the disorder amplitude is 0.1~meV (as compared with 0.01~meV in the main text). 
When disorder is present, the patterns are not as regular as in a quasi-ideal case. One can nevertheless discriminate a symmetric and an anti-symmetric pattern using the standard deviation of the X coordinate of the solitons mass center $X_c(y)$. This deviation reads as $\left<\Delta^2\right>=\left< (X_c(y) - \left< X_c(y) \right>_y)^2 \right>_y$, where $\left< \right>_y$ means averaging over the $Y$ coordinate. For symmetric patterns, the center is nearly aligned vertically, while for the anti-symmetric patterns the center undergoes sinusoidal trajectory and thus the deviation is larger. Computing or measuring experimentally this quantity allows to clearly distinguish different phases, as shown in the insets of Fig.~\ref{figpdd}. 
The limit cycle with oscillating solitons can still persist in the presence of disorder. This motion results in their broadening and "shallowing" as  clearly seen in the right inset of  Fig.~\ref{figpdd}. Importantly, the disorder can also suppress the motion of the solitons and thus the limit cycle phase region becomes much smaller with stronger disorder.
The diagram was calculated with randomly generated noise for each pair ($S, P$), which enhances the sampling.


\section{Solitons under $\pi$ phase shift}

\begin{figure*}[tbp]
  \centering
  \includegraphics[width=1.0\textwidth]{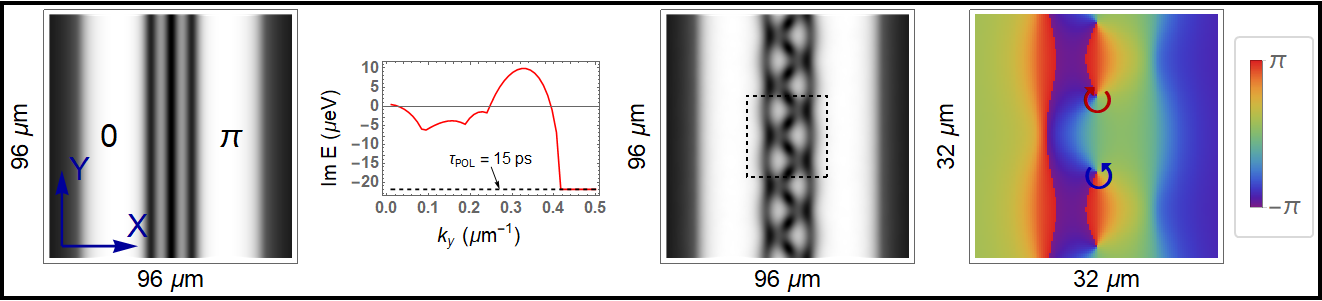}\\
  \includegraphics[width=1.0\textwidth]{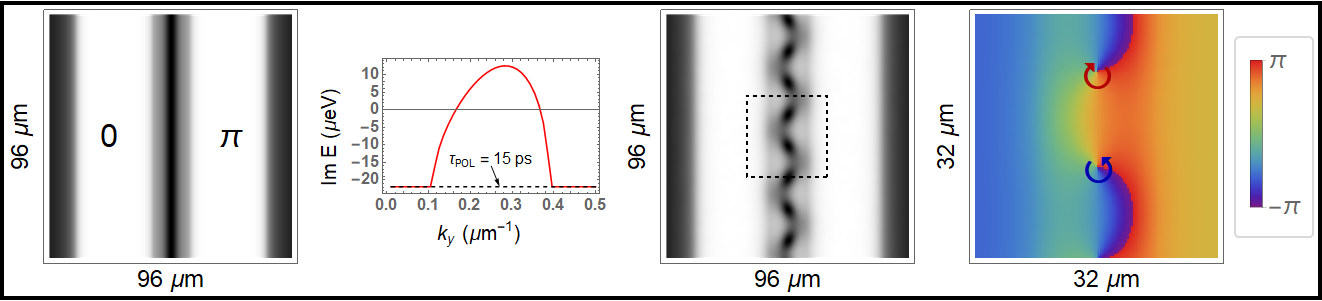}\\
  \caption{Modulational instability of guided solitons with a $\pi$ phase shift between the walls. $L=15$~$\mu$m, $P=0.8,1.2S_c$. (top, bottom). Columns: 1) stationary solution with even number of solitons between the high-density walls. 2) imaginary part of the energy of weak excitations of the stationary solution from the 1st column as a function of $k_y$. 3,4): stationary solution after the development of the modulational instability (density, phase)  in the presence of weak disorder.} \label{figpi}
\end{figure*}

In the main text, we show the results obtained when the phase of the pumping laser is homogeneous in space, and only the density profile is varying (allowing to obtain the high-density walls). In this subsection, we present additional results concerning the formation of solitons and their stability for a $\pi$ phase difference between the pump at the walls. In this case, no support is used (otherwise it would exhibit different interference with the two pumping lasers of different phase). 

Figure~\ref{figpi} shows the results obtained in this configuration, with top and bottom rows corresponding to two different values of the pump $P$. The first column shows the spatial density profile with 3 or 1 solitons, depending on the distance available for them because of the pump-induced broadening of the walls. The second column shows the imaginary part of the energy obtained from the Bogoliubov-de Gennes analysis described above. Both curves exhibit a maximum with positive imaginary part, confirming the existence of modulational instability. The final stage of the development of this instability is shown in the 3rd column: it exhibits either 3 vortex chains or a single vortex chain. Finally, the phase distribution shown in the 4th column confirms the formation of vortices and anti-vortices evidenced by the density shown in the 3rd column.

\section{Solitons in a maze}

In Figs.~1, 2, 3 of the main text, we were considering a system infinite in the $Y$ direction (implemented by periodic boundary conditions). As soon as we consider a dead end instead of an infinite channel (see Fig. 4(c) of the main text), the "head" of the pair of solitons comes into play. This head can also be considered as a vertical domain wall between the domains. This wall is between a high-density region (the wall limiting the dead end) and a low-density region (the region with solitons). There are now 3 domain walls in the system: two of them (horizontal) are opposite to each other, whereas one (vertical) can propagate, depending on the conditions. The extra kinetic energy appearing because of the variation of the wavefunction in the transverse direction (vertical), absent in the infinite system (for a very large channel), changes the conditions of the local bistability loop. The flows toward the low density region are also larger and both effects together (kinetic energy and flows) make the low-density regime impossible for the same parameters for which it was possible in the infinite system. The domain wall starts to propagate, leading to the expansion of the high-density region, erasing progressively the solitons in the channel.

\begin{figure}
  \centering
  \includegraphics[width=1.0\linewidth]{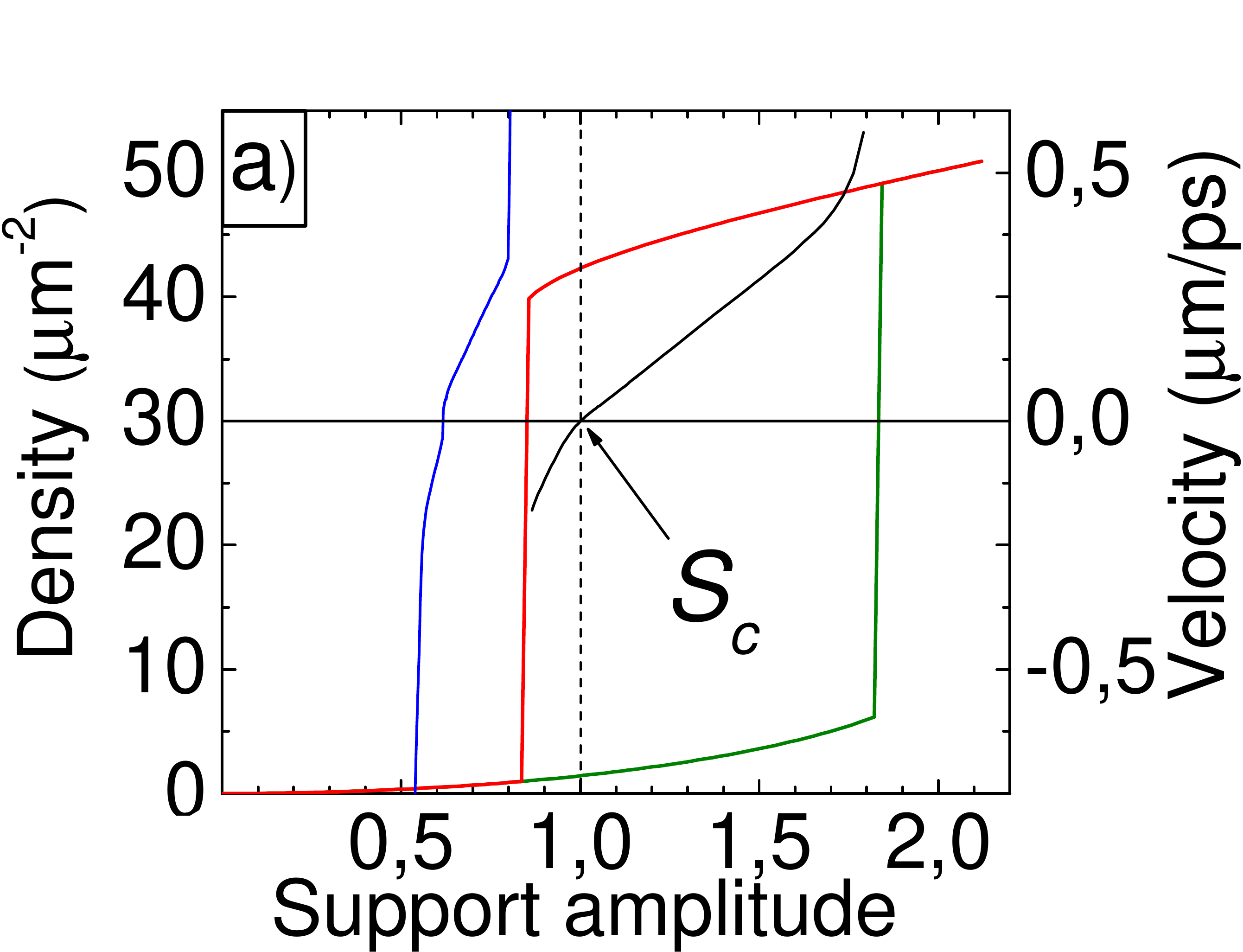}
  \caption{Velocity of DW in the 24 $\mu$m channel (blue curve) together with free DW velocity. Maze solving regime is realized in the velocity range from 0 to $\approx 0.2 \mu$m/ps. The bistability curve of a homogeneous system is plotted for a reference.} \label{figvelhead}
\end{figure}

Figure~\ref{figvelhead} plots the velocity of the "soliton head" (blue line) as a function of the Support in the 24 $\mu$m channel for the same value of the pump which was used in the  maze solving regime (see Fig. 3). One sees a behavior very similar to the "pure" domain wall from Fig. 1 (replotted as a black line), but shifted to much smaller values of $S$. This shift is due to the confinement of the domain wall in the transverse direction (vertical) and to the extra flows from the walls of the channel (shown in Fig.~4(c) of the main text).

\begin{figure}[tbp]
  \centering
  \includegraphics[width=0.99\linewidth]{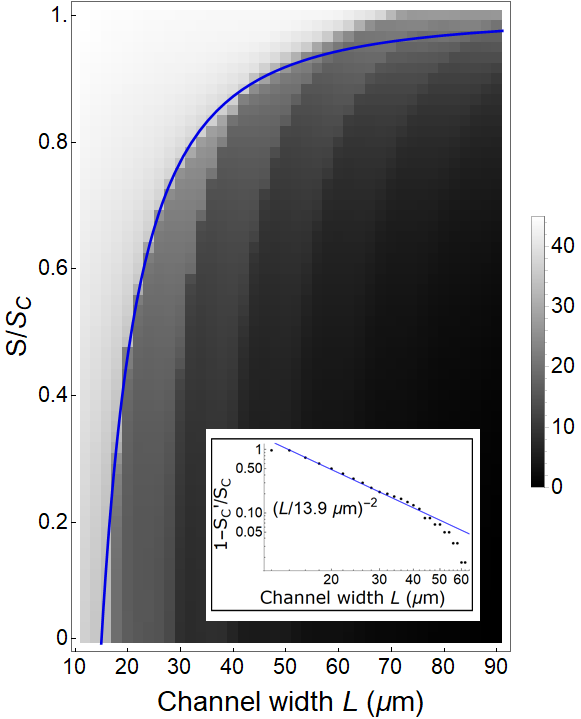}
  \caption{Mean polariton density in the corridor as a function of corridor width and support level $S$. White area in upper left corner corresponds to full filling the corridor with polaritons (repulsion of a dead end). Blue curve shows the inverse square fit $S/S_c=1-L_0^2/L^2$. The extracted values of $1-S/S_c$ and their fit for $L<30~\mu$m in log-log scale are shown in the inset.
  } \label{figpd2}
\end{figure}

To elucidate the contribution of these two effects, we plot an additional phase diagram (Fig.~\ref{figpd2}), showing for a single channel having a dead end the boundary between the two cases (solitons or high density in the channel) versus the channel width $L$ and the support amplitude $S$. The boundary between the two cases in shown in Fig.~4(d) of the main text. It is well described by an inverse square function with a single-parameter fit:
\begin{equation}
    \frac{S}{S_c}=1-\frac{L_0^2}{L^2}
\end{equation}
giving $L_0=13.9\pm0.1$~$\mu$m. This is the narrowest channel width where a pair of solitons can exist without support. The inverse square dependence is typical for a quantity depending on the confinement energy, and the value of $L\approx 14$~$\mu$m corresponds to the minimal width of a finite quantum well (with its depth of $0.14$~meV given by the laser detuning) where the two-node state can be confined for such particle mass (accounting for the finite broadening due to the lifetime). For a wider channel, the kinetic energy of the two-node state is lower, and the  support and the flows shift it upwards via the interactions. The inverse square law holds well up to $L=30$~$\mu$m, which is larger than the channel width used in the main text ($25$~$\mu$m). For even larger channels, the dependence becomes more complicated, because the low-density solution can contain more than 2 solitons, depending on $L$. The detailed study of this dependence is beyond the scope of the present work.

\begin{figure}[tbp]
  \centering
  \includegraphics[width=0.99\linewidth]{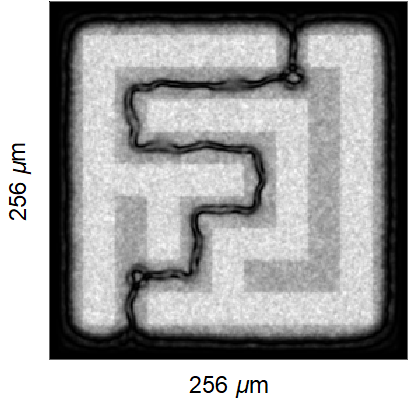}
  \caption{The maze solved at disorder of 0.1 meV. The other parameters have been increased with respect to the one used in the main text, namely detuning 3.2 times larger (0.45 meV), lifetime 3.2 times smaller (4.7 ps), and laser intensity 7 times larger, support $S\approx 5 S_c$ and channel walls $P+S \approx 16 S_c$.} \label{figdissolve}
\end{figure}

One could imagine that the motion of the soliton heads should lead to the switching of the whole maze structure to the high-density regime. However, this does not occur, because when the domain wall associated with a soliton edge disappearing from a dead end arrives to a T-junction where there is another soliton, the situation becomes fully equivalent to Fig.~2 of the main text: the T-junction with one channel in the high-density regime is actually not a T-junction any more, but just a channel with homogeneous walls (red curve in Fig.~4(d)). We conclude that the solitons can only disappear from the dead ends, whereas solitons connecting the exits of the maze persist, thus allowing to solve the maze. 

In the main text, we also plot the threshold support intensity versus the channel width for an open-end system, which exhibits higher values than for the dead end case as previously explained. Interestingly, our simulations show that in a large open-end channel the threshold support can even be larger than $S_c$. Indeed, in such case the lateral walls start to propagate towards the center where a 4-soliton structure forms. This structure is quite stable and produces a quantum pressure, which counteracts the pressure of the interactions and blocks the motion of the walls even if $S$ is slightly larger than $S_c$. We remind that  the fact that the threshold for the open end channels is higher than for a dead end channel, is what allows the head domain wall to withdraw from from the dead ends, but to stop at the crossroads, and therefore the maze solving process.

\subsection{Effects of disorder on the maze solving property}
As discussed in the main text, in presence of a large disorder the detuning has to be increased.
Fig.~\ref{figdissolve} is computed for a realistic disorder of 0.1 meV. To avoid disorder disturbance, we modified all parameters (detuning, lifetime, pump and support intensity), in order for the kinetic and interaction energies to be at least 3 times larger than the disorder amplitude. One should note that the used parameters are perfectly accessible experimentally. As one can see, the maze remains well defined and is solved, despite the disorder value.

\section{Supplementary Movies}

The supplementary video file \textsf{video.avi} shows an example of the solution of a large maze with time. The image size is $1024~\mu$m$\times 1024~\mu$m. The maze solving time is 16~ns. We see that the vortex streets withdraw from the dead ends and only the vortex street connecting the two exits of the maze remains.

An additional movie \textsf{video2.avi} shows the solution of a small maze (similar to that of Fig.~4 of the main text) over time, but with a large disorder amplitude, as shown in Fig.~\ref{figdissolve}.

\end{document}